\documentclass[12pt,a4paper,dvips]{article}
\usepackage{a4p}
\usepackage{cite,mcite}
\usepackage{graphicx}
\usepackage{epsfig}
\usepackage{amssymb,amsmath}
\usepackage{l3_title,ifthen}
\usepackage{physics,Lep}
\usepackage{a4p}

\date{September 20, 1999}
\preprint{99-129}
\newlength{\capindent}
\newlength{\capwidth}
\newlength{\figwidth}
\newcommand{\icaption}[2][!*!,!]{\hspace*{\capindent}%
  \begin{minipage}{\capwidth}
    \ifthenelse{\equal{#1}{!*!,!}}%
      {\caption{#2}}%
      {\caption[#1]{#2}}
  \end{minipage}}

\begin{document}

\begin{titlepage}

  \title{Single and Multi-Photon Events with Missing Energy \\
    in {\boldmath \epem} Collisions at {\boldmath \rts\ = 189 \gev}}

  \author{The L3 Collaboration}

%
%
  \begin{abstract}
    Single and multi-photon events with missing energy are analysed
    using data collected with the L3 detector at LEP at a
    centre-of-mass energy of 189 \gev, for a total of 176~\pb\  of
    integrated luminosity.
    The cross section of the process \epem\ \ra\ \nnbar\gam(\gam) is
    measured and the number of light neutrino flavours is determined
    to be \( N_\nu = 3.011 \pm 0.077 \) including lower energy
    data. Upper limits on cross sections of supersymmetric processes
    are set and interpretations in supersymmetric models provide
    improved limits on the masses of the lightest neutralino and the
    gravitino.  Graviton-photon production in low scale gravity models
    with extra dimensions is searched for and limits on the energy
    scale of the model are set exceeding 1 \tev{}  for two extra
    dimensions.  \end{abstract} 

\submitted

\end{titlepage}

\def\DELTAM{\ensuremath{\Delta m}}%
\def\Msel{\ensuremath{m_{\tilde\e_\mathrm{L}}}}%
\def\Mser{\ensuremath{m_{\tilde\e_\mathrm{R}}}}%
\def\Mselr{\ensuremath{m_{\tilde\e_\mathrm{L,R}}}}%
\def\Mchi{\ensuremath{m_{\tilde\chi^0_1}}}%
\def\Mchii{\ensuremath{m_{\tilde\chi^0_2}}}%
\def\Mcha{\ensuremath{m_{\tilde\chi^\pm_1}}}%
\def\Msnu{\ensuremath{m_{\tilde\nu}}}%
\def\Msml{\ensuremath{m_{\tilde\mu_\mathrm{L}}}}%
\def\Mstat{\ensuremath{m_{\tilde\tau_2}}}%
\def\MP{\ensuremath{m_{\mathrm{P}}}}%
\def\MG{\ensuremath{m_{\gravin}}}%
\def\susyl#1{\ensuremath{\tilde{#1}_\mathrm{L}}}%
\def\susyr#1{\ensuremath{\tilde{#1}_\mathrm{R}}}%
\def\susylr#1{\ensuremath{\tilde{#1}_\mathrm{L,R}}}%
\def\tanb{\ensuremath{\tan \beta}}%

%
%
\section{Introduction}

In the Standard Model \cite{standard_model} single or multi-photon
events with missing energy are produced via the reaction \epem\ \ra\
\nnbar\gam(\gam) which proceeds through $s$--channel Z exchange and
$t$--channel W exchange. Searches for single and multi-photon final
states as well as measurements of the \epem\ \ra\ \nnbar\gam(\gam)
cross section have already been performed by L3
\cite{papl3gg96,papl3gg97,papl3nu} and by other LEP experiments
\cite{paplepgg} at lower centre-of-mass energies. For the first time,
the determination of the number of light neutrino species from single
photon events at energies above the Z resonance is reported here.

In supersymmetric models \cite{susy} different supersymmetry (SUSY)
breaking mechanisms lead to different phenomenologies. The SUSY
breaking scale, $\sqrt{F}$, or equivalently the gravitino mass (\( \MG
= F/[\sqrt{3/(8\pi)}\,\MP] \) where \MP\  is the Planck mass),
is considered as a free parameter.  Three different scenarios are
distinguished: heavy, light and superlight gravitinos.

In gravity-mediated SUSY breaking models (SUGRA) the gravitino is
heavy (\( 100 \gev \lesssim \MG \lesssim 1 \tev \)) and thus does not
play a role in production or decay processes. The lightest neutralino
is the lightest supersymmetric particle (LSP), which is stable under
the assumption of R--parity \cite{rparity} conservation and escapes
detection due to its weakly interacting nature. In this scenario,
single or multi-photon signatures arise from pair-production of
neutralinos (\chinon\chinonn\  and \chinonn\chinonn)
\cite{neuprod}. Subsequent one-loop decays of \chinonn\  into
\chinon\gam\  have a branching fraction close to 100\% if one of the
two neutralinos is pure photino and the other pure higgsino
\cite{neugam}.  In general, neutralinos are mixtures of photinos,
zinos and higgsinos.

In models with Gauge-Mediated SUSY Breaking (GMSB) \cite{gmsb}, a
light gravitino (\( 10^{-2}\ev \lesssim \MG \lesssim 10^2\ev \)) is
the LSP. In this case the gravitino plays a fundamental role in the
decay of SUSY particles. In particular, the \chinon\  is no longer
stable and decays through \chinon\ \ra\ \gravin\gam\  if it is the
next-to lightest supersymmetric particle (NLSP)
\cite{neugra}. Pair-production of the lightest neutralino leads to a
two-photon plus missing energy signature in the detector.

When the scale of local supersymmetry breaking is decoupled from the
breaking of global supersymmetry as in no-scale supergravity models
\cite{noscalesugra}, the gravitino can be superlight (\( 10^{-6}\ev
\lesssim \MG \lesssim 10^{-4}\ev \)). Then, it is produced not only in
SUSY particle decays but also directly in pairs \cite{zwirner} or
associated with a neutralino \cite{lnz}. Pair-production of gravitinos
accompanied by initial state radiation leads to a single photon
signature. This signature also arises in \chinon\gravin\  production
when the neutralino decays radiatively to gravitino and photon.

Recently, it has been proposed that the fundamental gravitational
scale in quantum gravity models with extra dimensions is as low as the
electroweak scale \cite{qgrav1} thus naturally solving the hierarchy
problem. Within the framework of these models real gravitons are
produced in \epem\  collisions through the process \epem \ra \gam G,
where the graviton escapes undetected leading to a single photon plus
missing energy signature.

%
%
\section{Data Sample and Simulation}

In this analysis we use the data collected by the L3 detector
\cite{L3-DETECTOR} during the high energy run of LEP in 1998
corresponding to an integrated luminosity of 176.4~\pb\  at an average
centre-of-mass energy of \rts\ = 188.6 \gev,  hereafter denoted 189 \gev.

Monte Carlo events for the following Standard Model processes are
simulated: \epem\ \ra\ \nnbar\gam(\gam) with {\tt KORALZ}
\cite{KORALZ}, \epem\ \ra\ \gam\gam(\gam) with {\tt GGG} \cite{GGG},
Bhabha scattering for large scattering angles with {\tt BHWIDE}
\cite{BHWIDE}, and for small scattering angles with {\tt TEEGG}
\cite{teegg}, and four-fermion final states specifically the processes
\epem\ \ra\ \epem\epem\  with {\tt DIAG36} \cite{DIAG}, and \epem\
\ra\ \epem\nnbar\  with {\tt EXCALIBUR} \cite{excalibur_new}.

SUSY processes are simulated with the Monte Carlo program {\tt
SUSYGEN} \cite{SUSYGEN2.2} for SUSY particle masses
($m_\mathrm{SUSY}$) between zero and the kinematic limit and, in
\chinon\  LSP scenarios, for $\DELTAM=m_\mathrm{SUSY} -
m_\mathrm{LSP}$ between 1 \gev{}  and $m_\mathrm{SUSY}$. To ensure the
radiative decay of the neutralino, the scalar electron ($\susyr{\e}$)
mass is set to 100 \gev, except for \chinon\gravin\  production where
it is set to 200 \gev. The detector response is simulated using the
{\tt GEANT} program \cite{geant}, which takes into account the effects
of energy loss, multiple scattering and showering in the detector.

%
%
\section{Event Selection}

Electrons and photons are measured accurately by the BGO
electromagnetic calorimeter. They are required to have an energy
greater than 0.9 \gev. The shape of their energy deposition must be
consistent with an electromagnetic shower. Electrons are defined as
electromagnetic clusters matched with a charged track reconstructed in
the central tracking chamber. Identified conversion electrons from
photons interacting with the beam pipe or with the silicon microvertex
detector, 4\% of the total, are also accepted as photons. Bhabha
events and \epem\ \ra\ \gam\gam(\gam) events that are fully contained
in the calorimeter are used to check the particle identification as
well as the energy resolution, which is 1\% for high energy electrons
and photons in both the barrel and the endcaps. The barrel region is
defined as the polar angle range $43^\circ < \theta < 137^\circ$ with
respect to the beam axis and the endcap region as the polar angle
range $14^\circ < \theta < 36^\circ$ or $144^\circ < \theta <
166^\circ$.

\subsection{High Energy Photons}

The selection of high energy single and multi-photon events requires
at least one photon with energy greater than 5 \gev{}  in the barrel or
endcaps region.  There must be no charged tracks 
apart from those consistent with photon conversion. The following cuts
are imposed to suppress events which do not consist of photons only in
the final state. The energy not assigned to identified photons has to
be smaller than 10 \gev{}  and the energy measured in the
electromagnetic calorimeter between BGO barrel and endcaps must be
smaller than 7 \gev.  There must be no track in the muon chambers and
at most one BGO cluster not identified as a photon.

To reduce the background from radiative Bhabha events with particles
escaping along the beam pipe, as well as from the process \epem\ \ra\
\gam\gam(\gam), events with less than 20 \gev{}  transverse momentum are
rejected if energy is observed in the small polar angle detectors
covering an angular range of $1.5^\circ - 11^\circ$.
The total transverse momentum of photons is required to be greater
than 5 \gev{}  if no second photon with  energy greater than 5 \gev{}  is
found. If two calorimetric clusters are present and if only one is
identified as a photon, their acollinearity must be greater than
5.2$^\circ$ and their acoplanarity must be greater than
2.4$^\circ$. Furthermore, energy clusters in the hadron calorimeter
(HCAL) must have less than 3 \gev{}  energy if a photon is detected with
an acoplanarity less than 15$^\circ$ to the HCAL cluster.

When a second photon with energy above 5 \gev{}  is present, the total
transverse momentum must be greater than 3 \gev{}  and the recoil mass
must be larger than 20 \gev. If the total transverse momentum is
smaller than 30 \gev, the acollinearity is required to be larger than
8.1$^\circ$ and the acoplanarity to be larger than 5.2$^\circ$. If the
transverse momentum is smaller than 20 \gev, the missing momentum
direction is required to be at least 7$^\circ$ away from the beam
pipe. If the acoplanarity is smaller than 2.4$^\circ$, the recoil mass
must be greater than 50 \gev.

To suppress cosmic ray background, we require for photon energies
smaller than 15 \gev, that the most energetic photon is not aligned
with hits in the muon detector. For photon energies larger than
15 \gev, there must be at least one scintillator time measurement
within $\pm$5~ns of the beam crossing time. Furthermore, an event is
rejected if more than 20 hits are found in the central tracking
chamber in a 1~cm road between any pair of energy depositions in the
BGO.

\begin{figure}
  \begin{center}
    \includegraphics[width=0.68\linewidth]{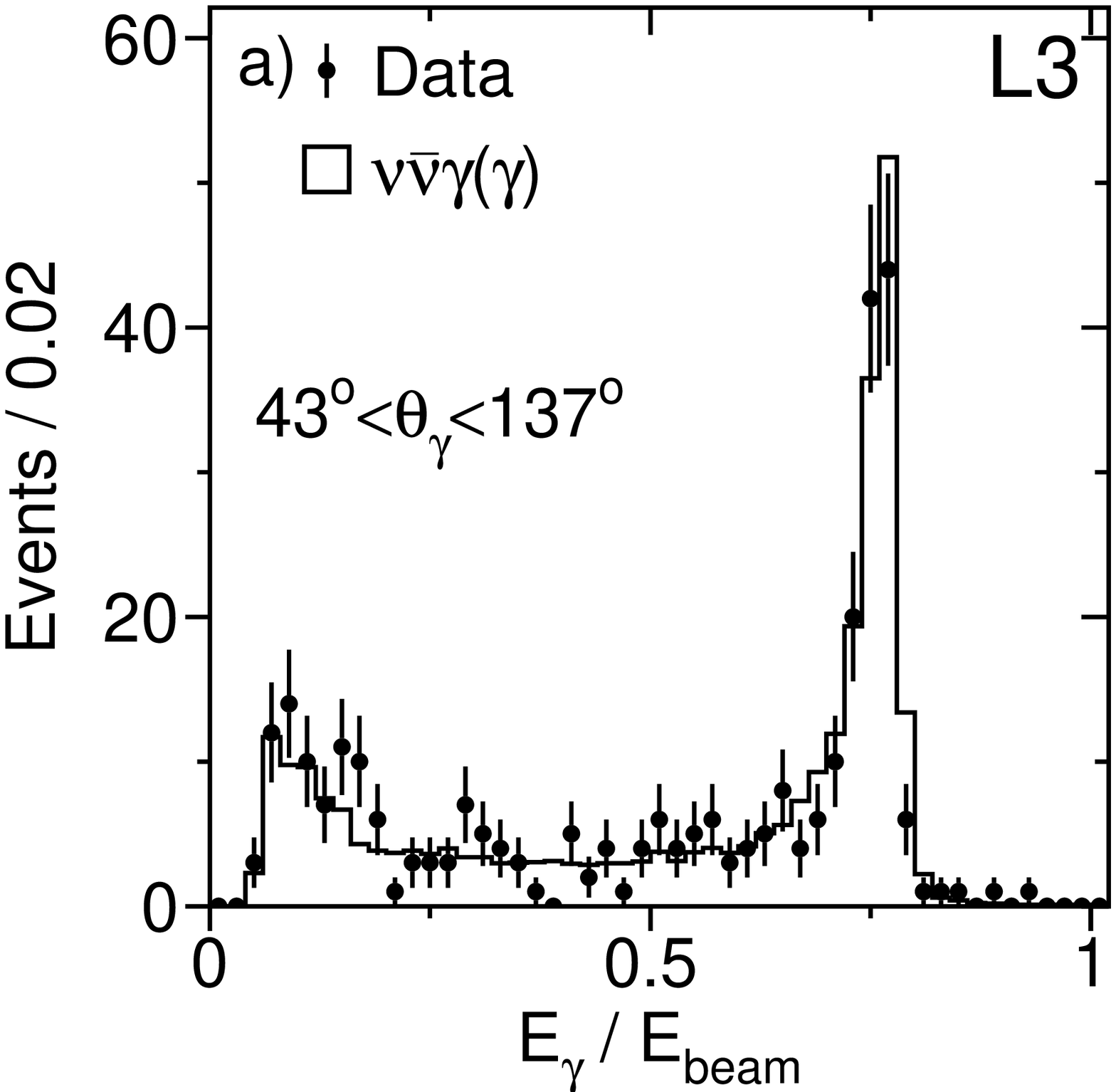}
    \includegraphics[width=0.68\linewidth]{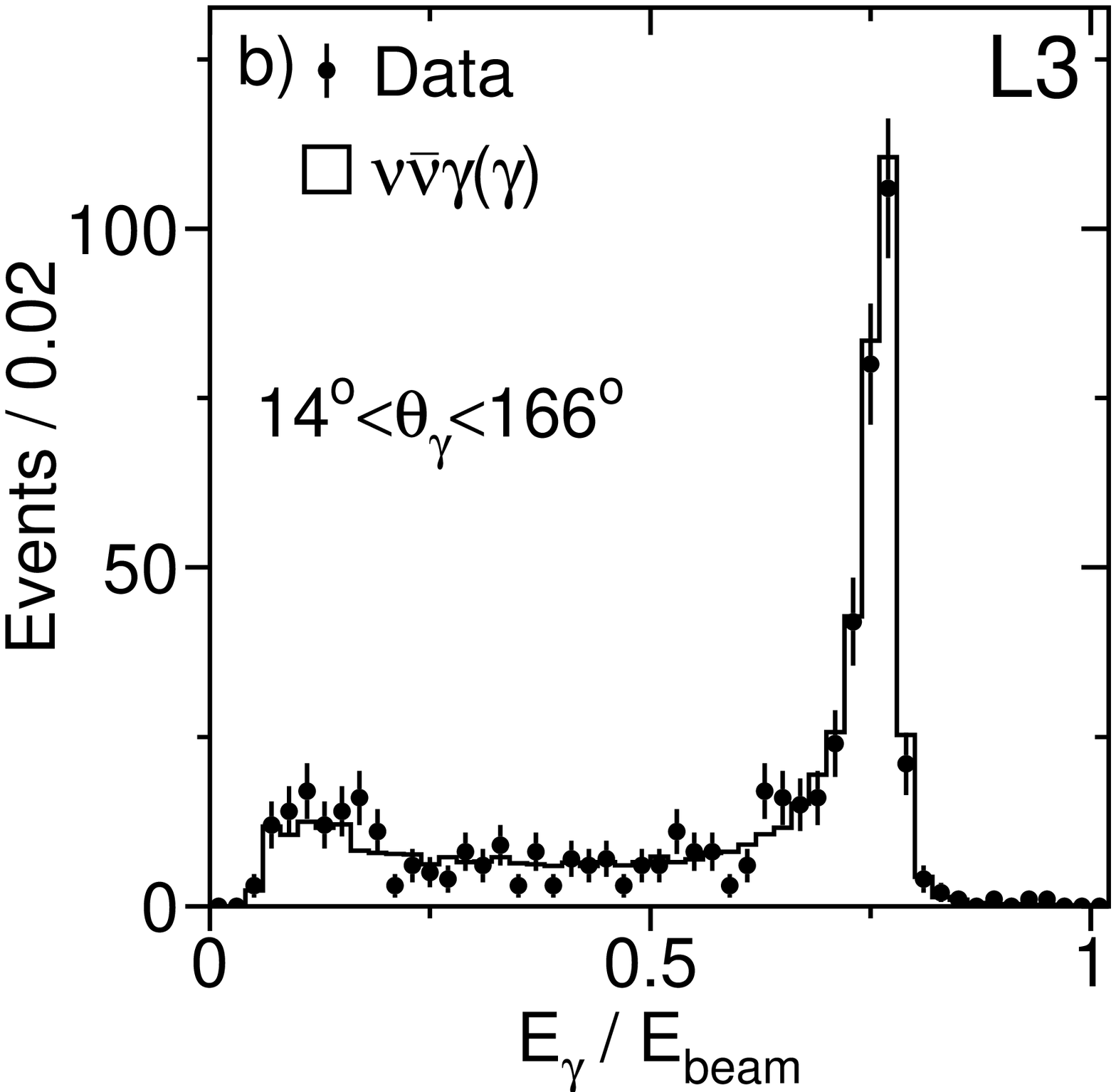}
    \icaption{\label{fig:egamma}(a) Energy of the highest energetic
      photon normalised to the beam energy for single and multi-photon
      events at 189 \gev{}  in the barrel region. (b) Same distribution
      with the endcaps included.}
  \end{center}
\end{figure}

The number of events with one or more photons is listed in table
\ref{tab:sgamsel} together with the predicted rates for
\nnbar\gam(\gam) and other processes originating from \epem\
collisions. The cosmic ray background in the event sample is estimated
from studies of out-of-time events and also listed in table
\ref{tab:sgamsel}. Figure \ref{fig:egamma} shows the energy spectrum
of the most energetic photon normalised to the beam energy for single
and multi-photon events.

\begin{figure}
  \begin{center}
    \includegraphics[width=0.68\linewidth]{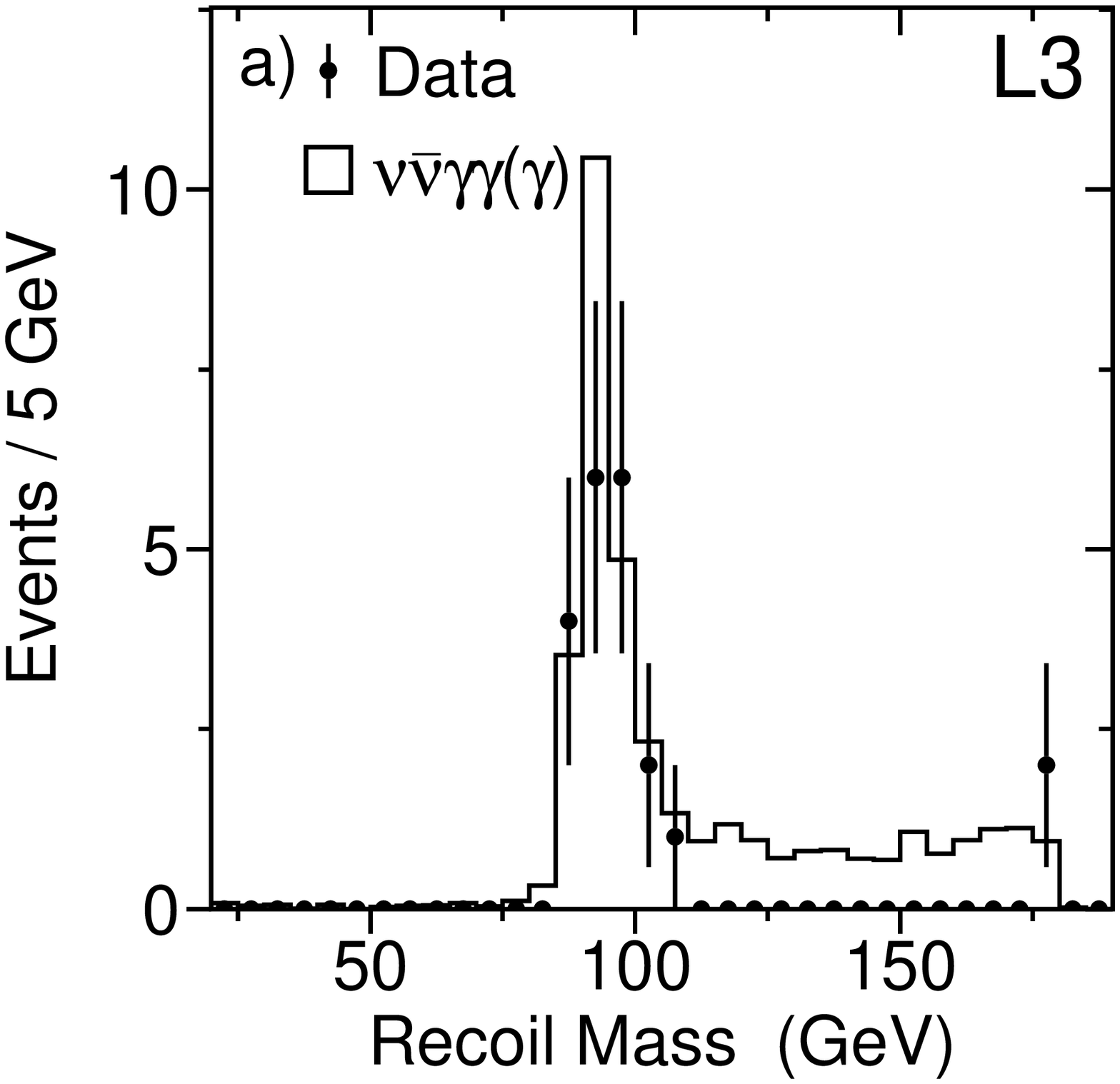}
    \includegraphics[width=0.68\linewidth]{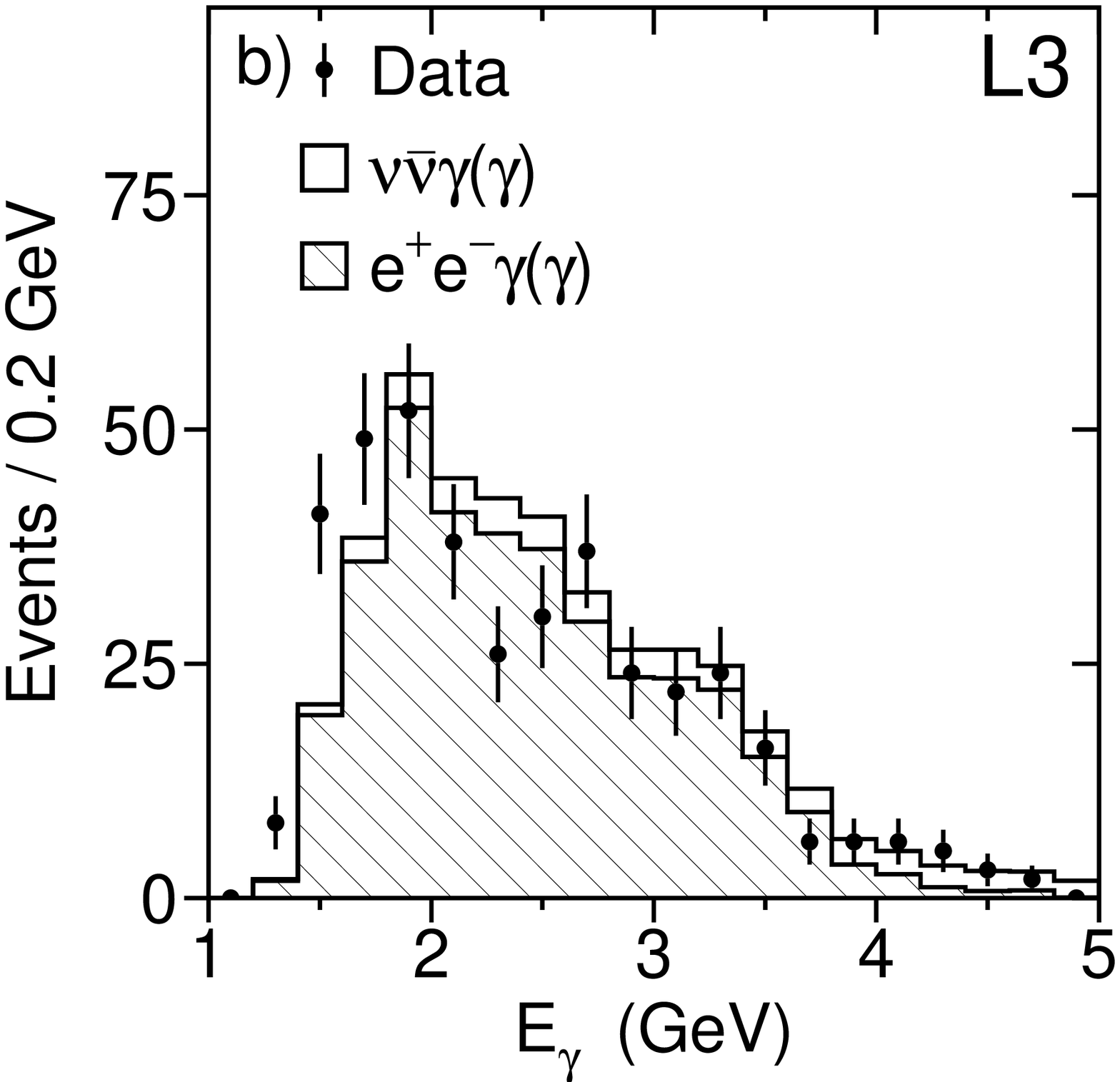}
    \icaption{\label{fig:recmegalo}(a) Recoil mass for the
      multi-photon sample. (b) Low energy part of the energy spectrum of
      single photon events.}
  \end{center}
\end{figure}

For the sub-sample of events with two or more photons a minimum energy
for the second photon of 1 \gev{}  is required. In the data 21 events
are observed compared to a Monte Carlo prediction of 36.2 events, see
table \ref{tab:sgamsel}. For recoil masses larger than 110 \gev{}  we
observe 2 events compared to an expectation of 12.7 events. Figure
\ref{fig:recmegalo}(a) shows the two-photon recoil mass distribution.
The lack of data compared to the Monte Carlo prediction has been
subject to extensive investigations concerning the performance of
sub-detectors and triggers used in this analysis. The noise level of
sub-detectors is studied using randomly triggered beam-gate
events. The performance of the electromagnetic calorimeter is
cross-checked with Bhabha events and events from \epem\ \ra\
\gam\gam(\gam). Triggers important for single and multi-photon events
are investigated using single electron\footnote{Radiative Bhabha
  scattering events where one electron and a photon have a very low
  polar angle, and only a low energetic electron is scattered with a
  large polar angle.}
and Bhabha events. The theoretical predictions for the cross section
of \epem\ \ra\ \nnbar\gam\gam(\gam) obtained from {\tt KORALZ} and
{\tt NUNUGPV} are found to agree within 5\%. No systematic effect is
found to explain the low two-photon rate. An independent analysis
leads to the same conclusion. Therefore the deficit is treated as a
statistical fluctuation.

\subsection{Low Energy Photons}

This selection extends the energy range for photons down to
1.3 \gev. It covers only the barrel region where a single photon
trigger is implemented with a threshold around 900 \mev{}
\cite{sgtrig}. There must be no other BGO clusters in barrel or
endcaps with more than 200 \mev. The energy in the HCAL must be less
than 6 \gev.  To reduce the rate of small angle Bhabha scattering no
energy deposit is allowed in the forward detectors. Events with a
track in the central tracking chamber
or in the muon chambers are rejected to reduce the rate of single
electron and cosmic muon events. To further reduce cosmic ray events
not pointing to the interaction region, cuts on the transverse shape
of the photon shower are made. For the simulation of the process
\epem\ \ra\ \epem\gam(\gam) the {\tt TEEGG} program is used which
includes fourth order contributions. In order to use this program, a
cut on the transverse momentum of the photon greater than 1.3 \gev{}  is
applied \cite{teegg}. From a study of single electron events its
precision is estimated to be at the 20\% level. The number of selected
events, predictions from \epem\  collision processes and an estimate
of cosmic ray contamination are listed in table
\ref{tab:sgamsel}. Figure \ref{fig:recmegalo}(b) shows the observed
photon energy spectrum compared to Monte Carlo prediction.

%
%
\section{Neutrino Production}

To measure the cross section of the \epem\ \ra\ \nnbar\gam(\gam)
process we restrict the analysis to photon energies above 5 \gev{} (see
table \ref{tab:sgamsel}) to ensure a good signal to background ratio.

The overall efficiency for \epem\ \ra\ \nnbar\gam(\gam) events
satisfying the kinematic requirements $E_\gam > 5 \gev$ and
$|\!\cos\theta_\gam| < 0.97$ is 60.77\%.  This efficiency includes a
correction of (2.58 $\pm$ 0.18)\% due to cosmic ray veto requirements,
which is estimated by studying single electron events. A correction of
(0.67 $\pm$ 0.07)\% due to detector noise sources not properly
simulated such as that induced by beam halo in the forward detectors
is also included and is quantified using randomly triggered beam-gate
events. The systematic error on the efficiency is composed of the
errors on the two corrections and several other sources including an
error of 0.34\% due to photon identification, of 0.60\% due to an
uncertainty on the amount of converted photons, and 0.21\% due to
limited Monte Carlo statistics. The total systematic error on the
efficiency amounts to 0.75\%.  The error on the luminosity is 0.4~\pb\
and on the total background contamination the error is 2.1 events.

\begin{figure}
  \begin{center}
    \includegraphics[width=\linewidth]{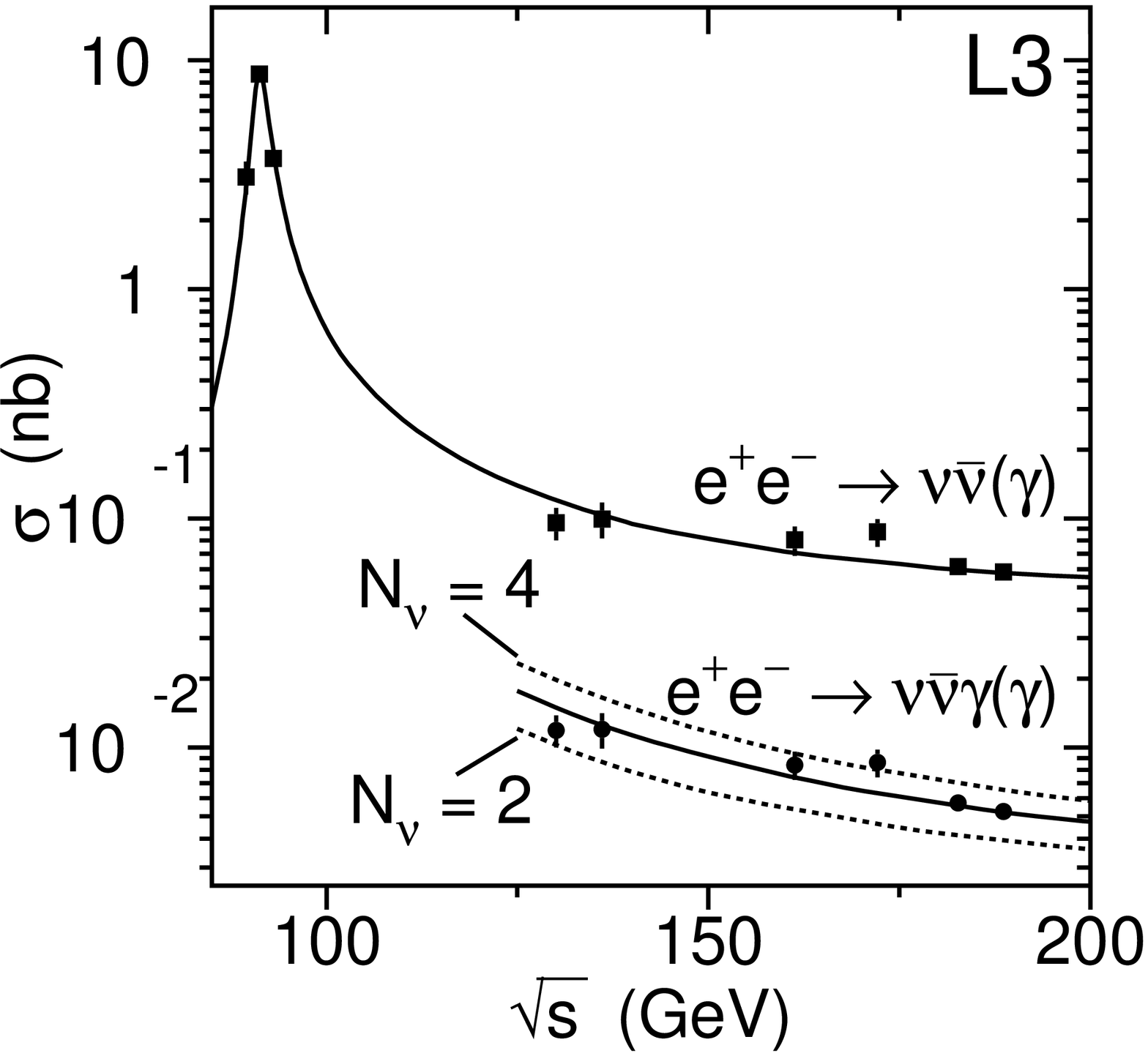}
    \icaption{\label{fig:nnxsection}Production cross section of \epem\
    \ra\ \nnbar(\gam) and \epem\ \ra\ \nnbar\gam(\gam) as a function
    of the centre-of-mass energy. Points with error bars represent the
    \nnbar\gam(\gam) measurements and squares with error bars are the
    extrapolation to \nnbar(\gam). The full line is the theoretical
    prediction for $N_\nu=3$ and dashed lines are predictions for
    $N_\nu = 2,4$ as indicated.}
  \end{center}
\end{figure}

The measured cross section at \rts\ = 188.6 \gev{}  is
\begin{equation*}
  \sigma_{\nnbar\gam(\gam)} = 5.25 \pm 0.22 \: \text{(stat)} \pm
  0.07 \: \text{(syst)}~\mathrm{pb}
\end{equation*}
to be compared to the prediction of the Standard Model of 5.28 $\pm$
0.05~pb obtained with {\tt KORALZ}, where the 1\% error accounts for
the theoretical uncertainty assigned to this process
\cite{NUNUGPV_NEW}. This measurement is extrapolated to a total cross
section for \epem\ \ra\ \nnbar(\gam) production of 58.3 $\pm$ 2.7~pb.
The prediction of the Standard Model obtained with {\tt KORALZ} is
58.6~pb. Figure \ref{fig:nnxsection} shows both the \nnbar\gam(\gam)
cross section measurement and the total neutrino-pair extrapolation
versus centre-of-mass energy together with the prediction of the
Standard Model and measurements at lower centre-of-mass energies.

To determine the number of light neutrino species a maximum likelihood
fit to the photon energy spectra is performed at each centre-of-mass
energy above the Z resonance. For each energy interval the theoretical
prediction is obtained by linearly interpolating {\tt KORALZ}
predictions for $N_\nu=1,2,3,4,5$. Due to the different contributions
to the energy spectrum from $\nu_\e\bar\nu_\e$ $t$--channel production
via W exchange and \nnbar\ $s$--channel production via Z exchange, this
method is more powerful than using the total cross section
measurement. In addition to the systematic error from the cross
section measurement, the theoretical uncertainty on the photon energy
spectrum -- estimated by comparing {\tt KORALZ} with {\tt NUNUGPV} --
is taken into account. The result is \( N_\nu = 3.05 \pm 0.11 \pm 0.04
\).

A compilation of the measurements at the different centre-of-mass
energies is shown in table \ref{tab:nonuav}. The precision of this
result is comparable with our previous measurement \cite{papl3nu} from
single photon events around the Z resonance. The combined measurement
is
\begin{equation*}
  N_\nu = 3.011 \pm 0.077 \, .
\end{equation*}
This result 
is more precise than the present world average on the number of light
neutrino families determined with the single photon method \cite{caso}.

%
%
\section{Limits on Supersymmetry}

The limits derived in the following are obtained from 189 \gev{}  data.
They correspond to a confidence level of 95\%.  Cross section limits
are calculated using a likelihood approach \cite{clopti} where the
spectra of a discriminant variable for data, background and signal
simulations are compared.

\subsection{Single Photon Signature}

For interpretations within SUSY frameworks the low energy photons are
included.  Here, the discriminant variable used to derive cross
section limits is the photon energy.

\begin{figure}
  \begin{center}
    \includegraphics[width=\linewidth]{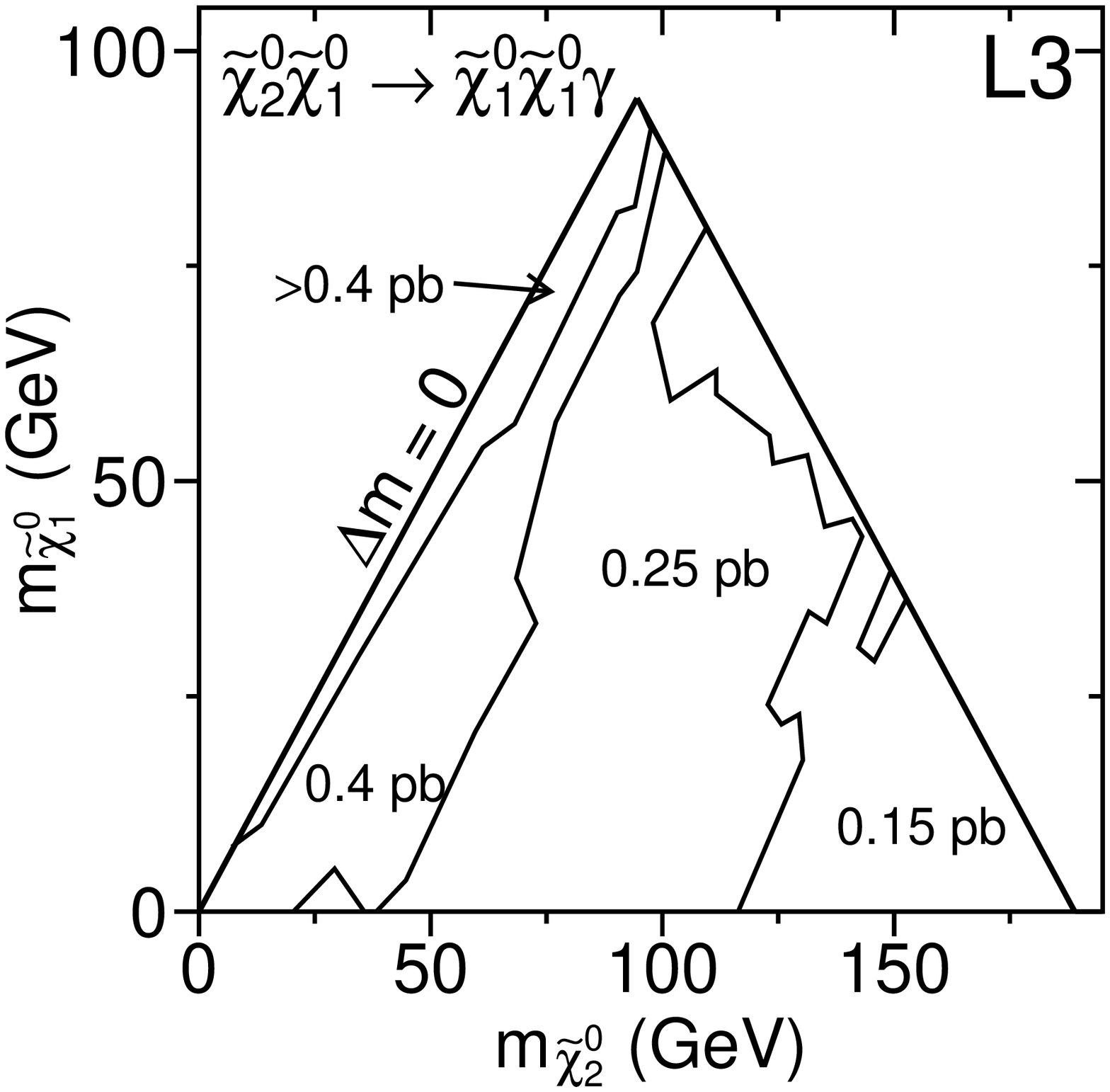}
    \icaption{\label{fig:n1ne2}Upper limits on the production cross
      section in picobarn for the process \epem\ \ra\ \chinonn\chinon\
      \ra\ \chinon\chinon\gam\  assuming 100\% branching ratio for
      \chinonn\ \ra\ \chinon\gam.}
  \end{center}
\end{figure}

In the heavy gravitino scenario, the single photon signature arises
from  the reaction \epem\ \ra\ \chinonn\chinon, which proceeds through
$s$--channel Z exchange and $t$--channel scalar electron exchange
(\susylr{\e}). Cross section upper limits shown in figure
\ref{fig:n1ne2} are set under the assumption of 100\% branching
fraction for \chinonn\ \ra\ \chinon\gam.
Typical detection efficiencies for this process are around 75\%.

\begin{figure}
  \begin{center}
    \includegraphics[width=0.66\linewidth]{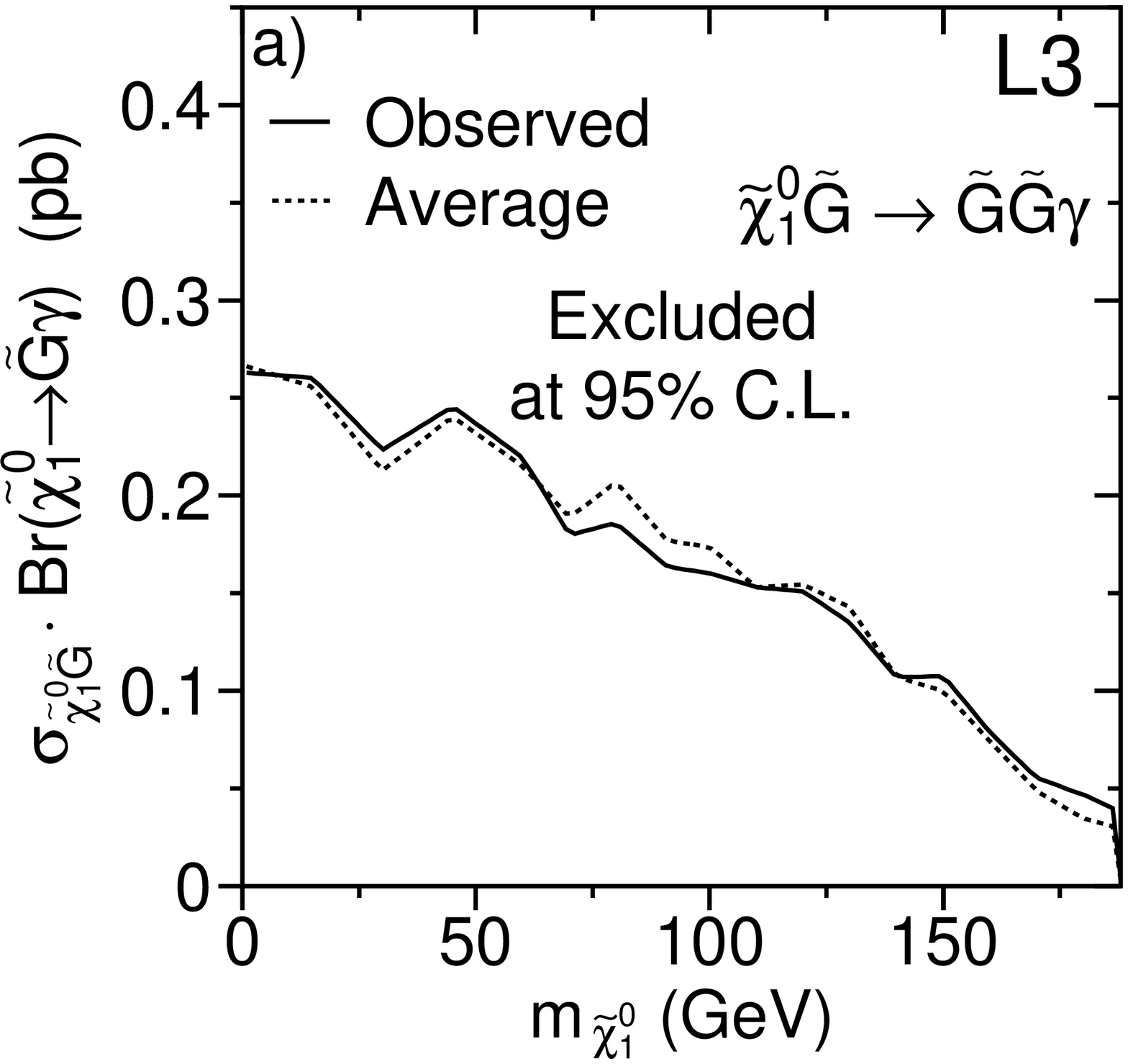}
    \includegraphics[width=0.66\linewidth]{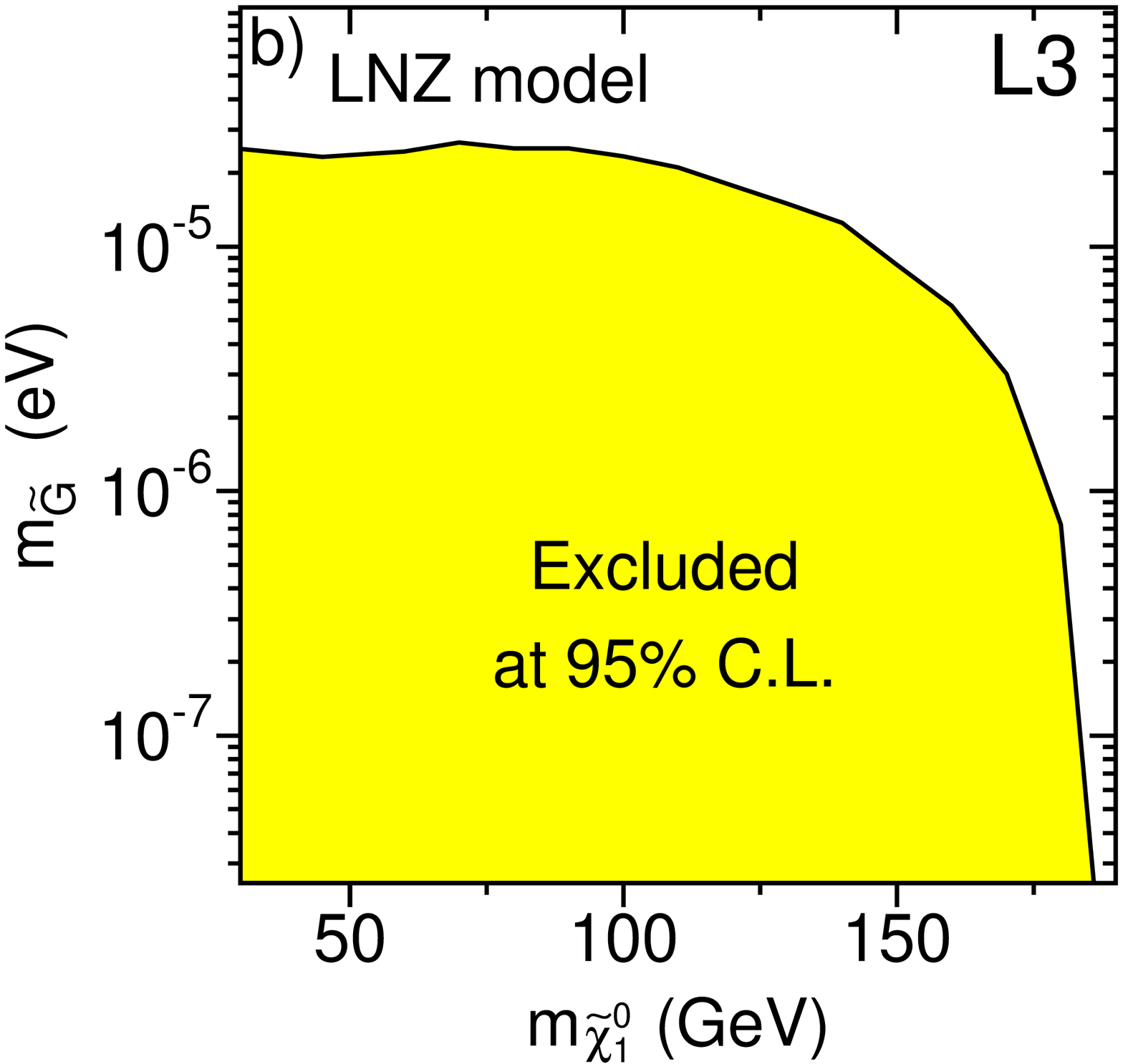}
    \icaption{\label{fig:n1gra}(a) Upper limits on the production
      cross section for the process \epem\ \ra\ \gravin\chinon\ \ra\
      \gravin\gravin\gam\  and average limit obtained using Monte Carlo
      trials with background only. (b) Region excluded in the LNZ
      model in the plane \MG\  versus \Mchi.}
  \end{center}
\end{figure}

Also the reaction \epem\ \ra\ \gravin\chinon\  proceeds through
$s$--channel Z exchange and $t$--channel \susylr{\e}
exchange. Efficiencies for this process with \chinon\ \ra\
\gravin\gam\  range between 64\% for \Mchi\ = 0.5 \gev{}  and 79\% for
\Mchi\  at the kinematic limit. The cross section upper limit as a
function of \Mchi\  is shown in figure \ref{fig:n1gra}(a) together
with the expected limit obtained in Monte Carlo trials with
background only.
The no-scale SUGRA model of \cite{lnz}, referred to as LNZ, has only
two free parameters -- gravitino and neutralino masses. The neutralino
is the NLSP, which is almost pure bino. Here, the dominant decay
channel is \chinon\ \ra\ \gravin\gam. The small contribution of the
decay into Z for \( \Mchi \gtrsim 100 \gev \) is taken into
account. Figure \ref{fig:n1gra}(b) shows exclusion contours in the
\MG\,--\,\Mchi\ plane.

If $m_\mathrm{NLSP} > \rts$, the process \epem\ \ra\ \gravin\gravin\
\cite{zwirner} is the only reaction to produce SUSY particles.
Accompanied by initial state radiation it leads to single or
multi-photon signatures. Following our analysis in \cite{papl3gg97} a
lower limit on the gravitino mass is derived
\begin{equation*}
  \MG > 8.9\cdot 10^{-6}\ev \, ,
\end{equation*}
corresponding to a lower limit on the SUSY breaking scale of $\sqrt{F}
> 192.3 \gev$.  The average lower limit for the gravitino mass obtained
in Monte Carlo trials with background only is $9.7\cdot
10^{-6}\ev$.

\subsection{Multi-Photon Signature}

\begin{figure}
  \begin{center}
    \includegraphics[width=\linewidth]{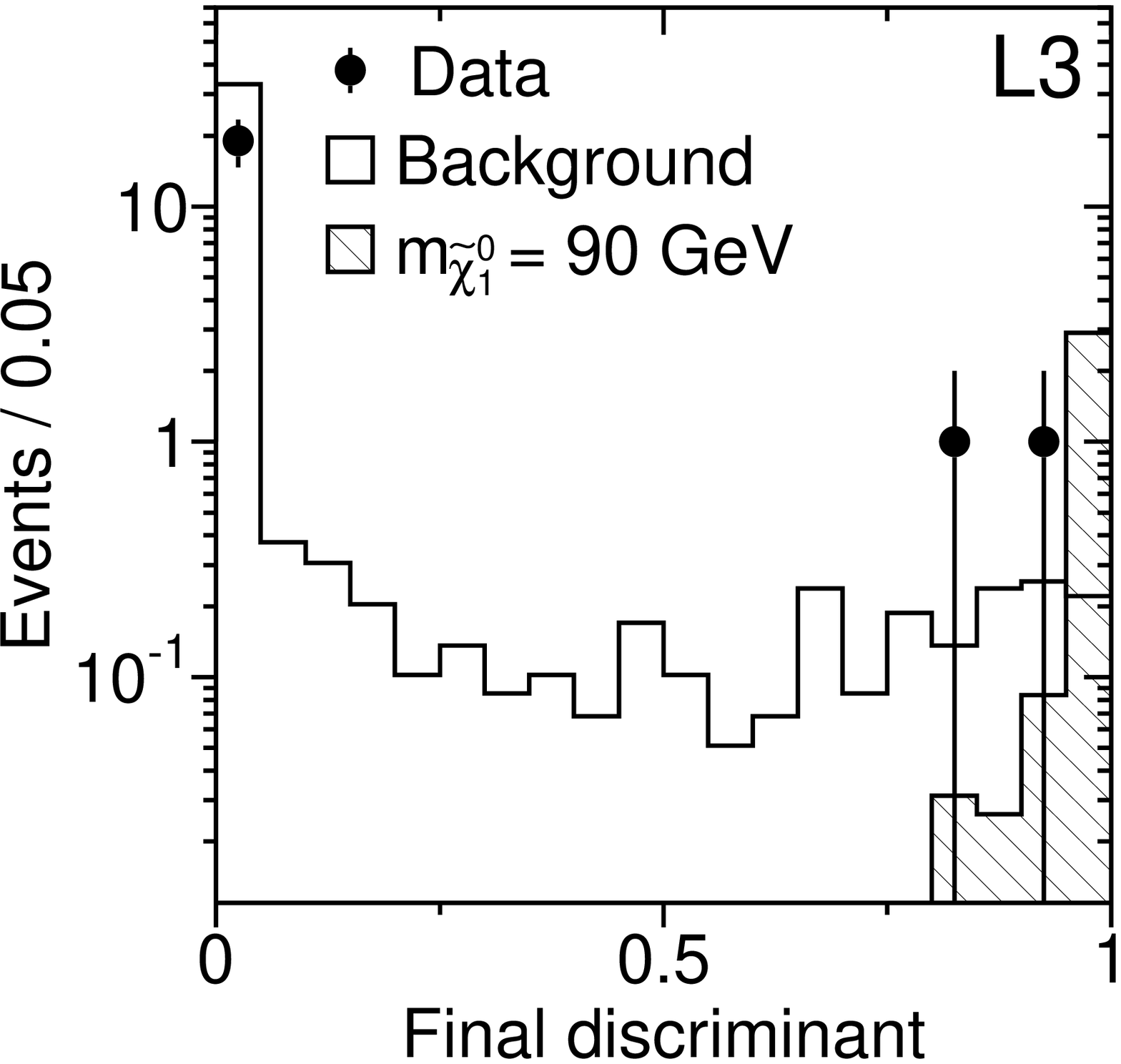}
    \icaption{\label{fig:discvar}Discriminant variable for
      \chinon\chinon\ \ra\ \gravin\gravin\gam\gam\  with \Mchi\ =
      90 \gev. The signal corresponds to the upper limit of 3.15 events
      derived for this mass point.}
  \end{center}
\end{figure}

Using a binned likelihood technique, the discriminant variable is
constructed for the multi-photon events combining the energies of the
two most energetic photons, their polar angles, recoil mass, and the
polar angle of the missing momentum vector. For each event class,
background and signal Monte Carlo processes, denoted by $j$, and each
input quantity $i$, a probability density function $f^i_j$ is computed
and the discriminant variable of an event is then given by
\begin{equation*}
  F(\vec{x}) = \frac{\prod_i p^i_\mathrm{signal}(x_i)}{\sum_j
    \prod_i p^i_j(x_i)} \quad \text{with} \quad p^i_j(x_i) =
  \frac{f^i_j(x_i)}{\sum_k f^i_k(x_i)} \; ,
\end{equation*}
where $x_i$ are the measured values of the six input variables of an
event. The distribution of the discriminant is shown in figure
\ref{fig:discvar} for \chinon\chinon\ \ra\ \gravin\gravin\gam\gam\
with \Mchi\ = 90 \gev.
The discrepancy between measurement and Standard Model prediction is
located in the background and not in the signal region. This holds
also for the other mass points and for the heavy gravitino scenario.

\begin{figure}
  \begin{center}
    \includegraphics[width=0.64\linewidth]{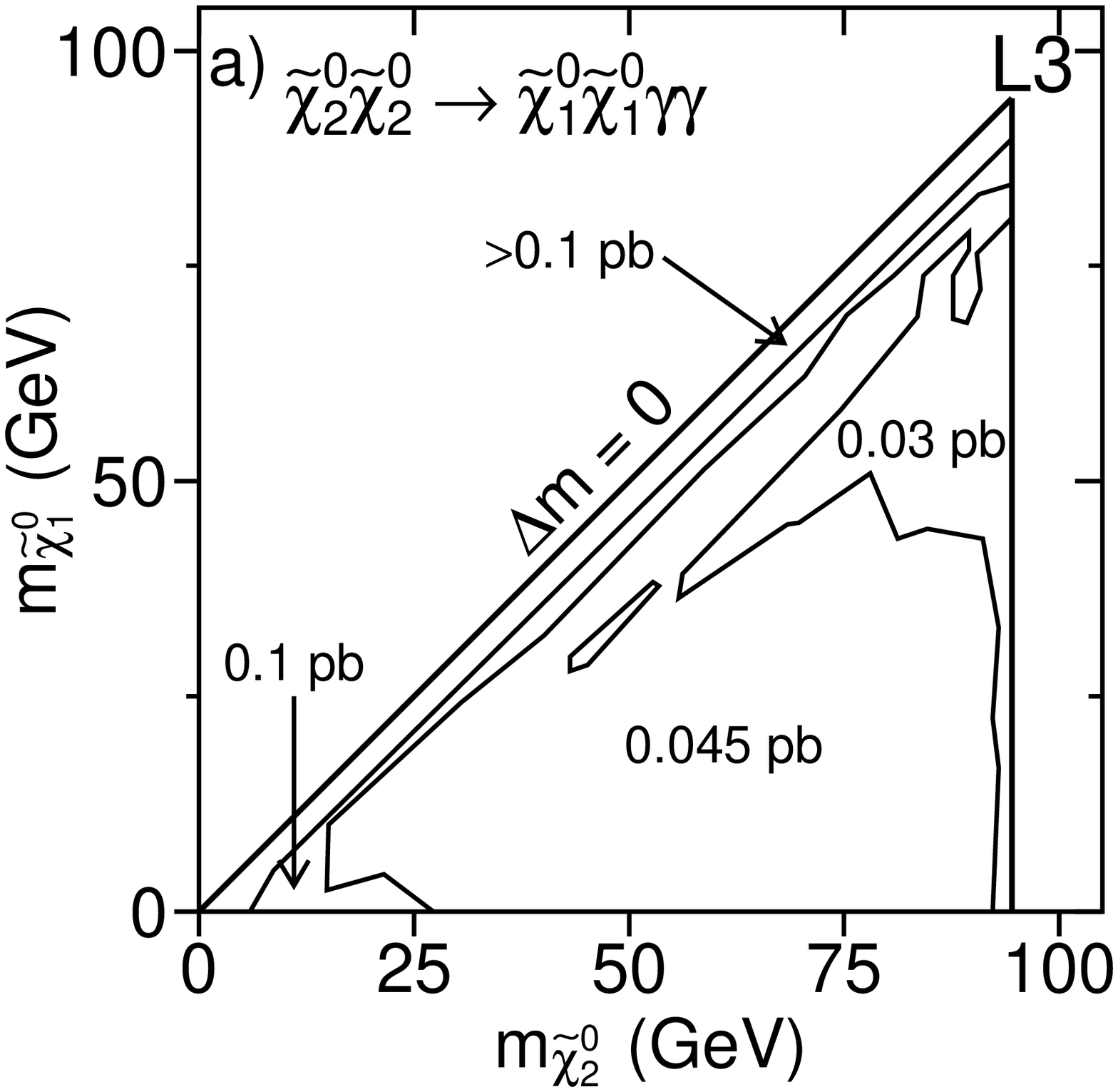}
    \includegraphics[width=0.64\linewidth]{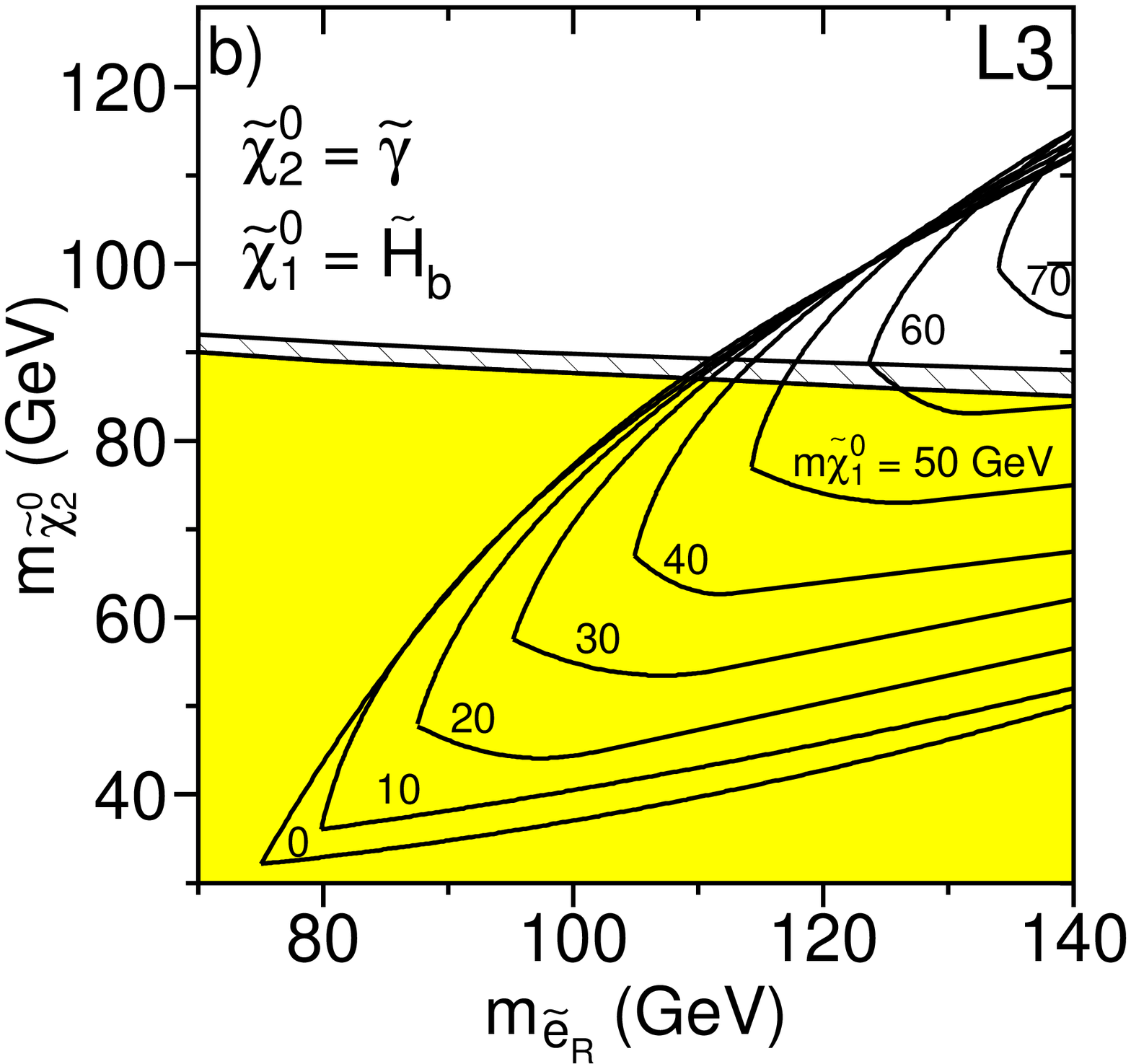}
    \icaption{\label{fig:n2ne2}(a) Upper limits on the production
      cross section in picobarn for the process \epem\ \ra\
      \chinonn\chinonn\ \ra\ \chinon\chinon\gam\gam. (b) Excluded region
      in the neutralino selectron mass plane. The shaded region
      corresponds to \Msel\ $\gg$ \Mser\  and the hatched region is
      additionally excluded when \Msel\ = \Mser. Regions kinematically
      allowed for the CDF event \protect\cite{cdfinterp12} as a function
      of \Mchi\  are indicated, where $ \chinon = \susy{H}_b =
      \susy{H}_1^0 \sin\beta + \susy{H}_2^0 \cos\beta $.}
  \end{center}
\end{figure}

In the heavy gravitino scenario, a two-photon signature is produced by
the process \epem\ \ra\ \chinonn\chinonn\  and subsequent decay to
\chinon\gam. Typical efficiencies for this process are around
64\%. Cross section upper limits are obtained as shown in figure
\ref{fig:n2ne2}(a).
The interpretation of the \e\e\gam\gam\  event with large transverse
missing energy observed by CDF \cite{cdfevent2} suggests a high
branching ratio for the radiative decay of the \chinonn\  in the heavy
gravitino scenario, which can be achieved if \chinonn\  is a pure
photino and \chinon\  is a pure higgsino. With this assumption, the
lower mass limit of \chinonn\  as a function of the scalar electron
mass is calculated for mass differences between \chinonn\  and
\chinon\  greater than 10 \gev{}  as shown in figure
\ref{fig:n2ne2}(b). For each \chinonn\  mass, the exclusion is
obtained using the most conservative cross section upper limit for any
$\DELTAM > 10 \gev$. The regions kinematically allowed
\cite{cdfinterp12} for the CDF event are overlayed in figure
\ref{fig:n2ne2}(b). The two exclusions obtained for equal masses of
\susylr{\e} and for \( \Msel \gg \Mser \) are shown in the interesting
mass range for \Mser.

\begin{figure}
  \begin{center}
    \includegraphics[width=0.65\linewidth]{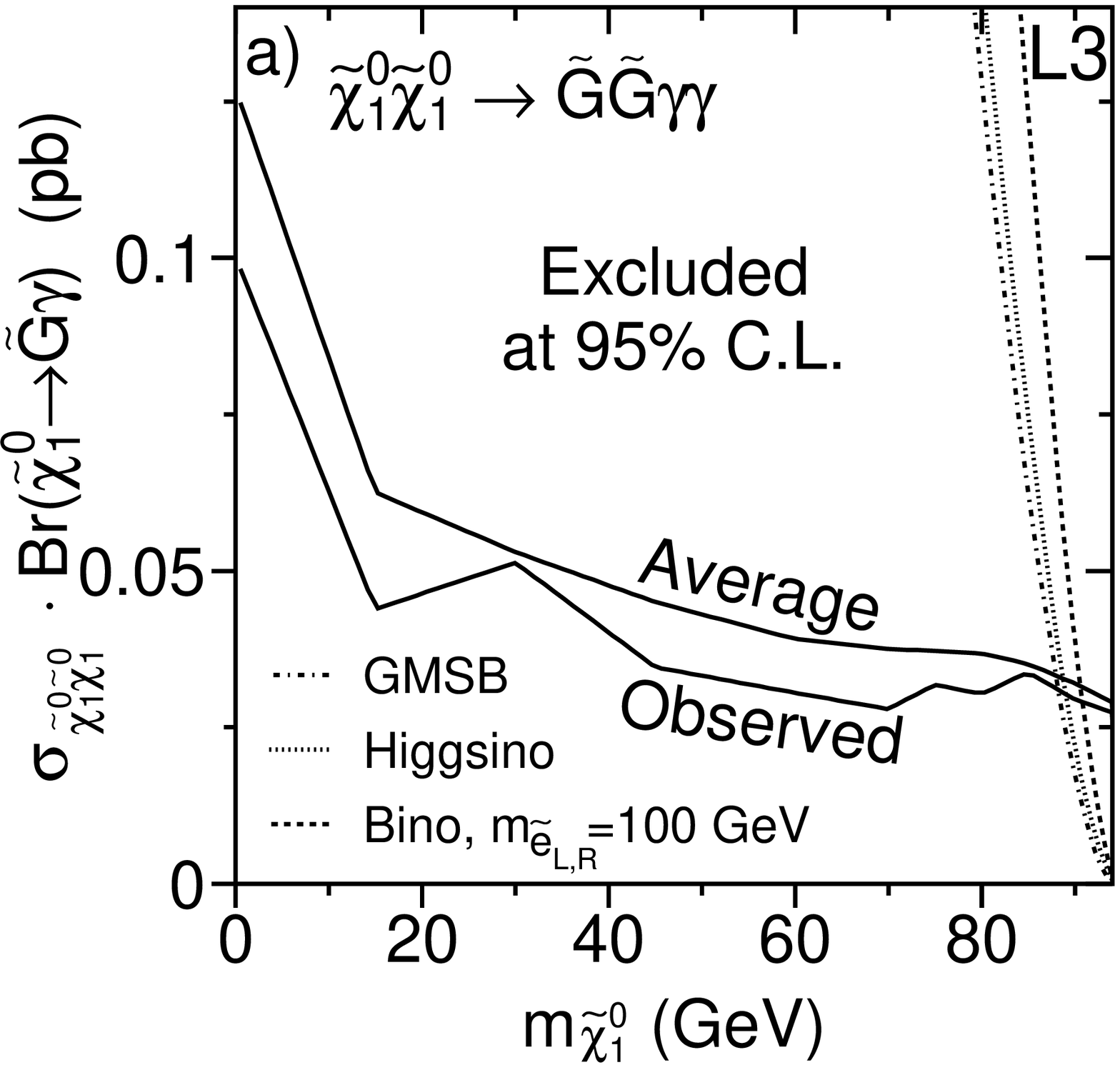}
    \includegraphics[width=0.65\linewidth]{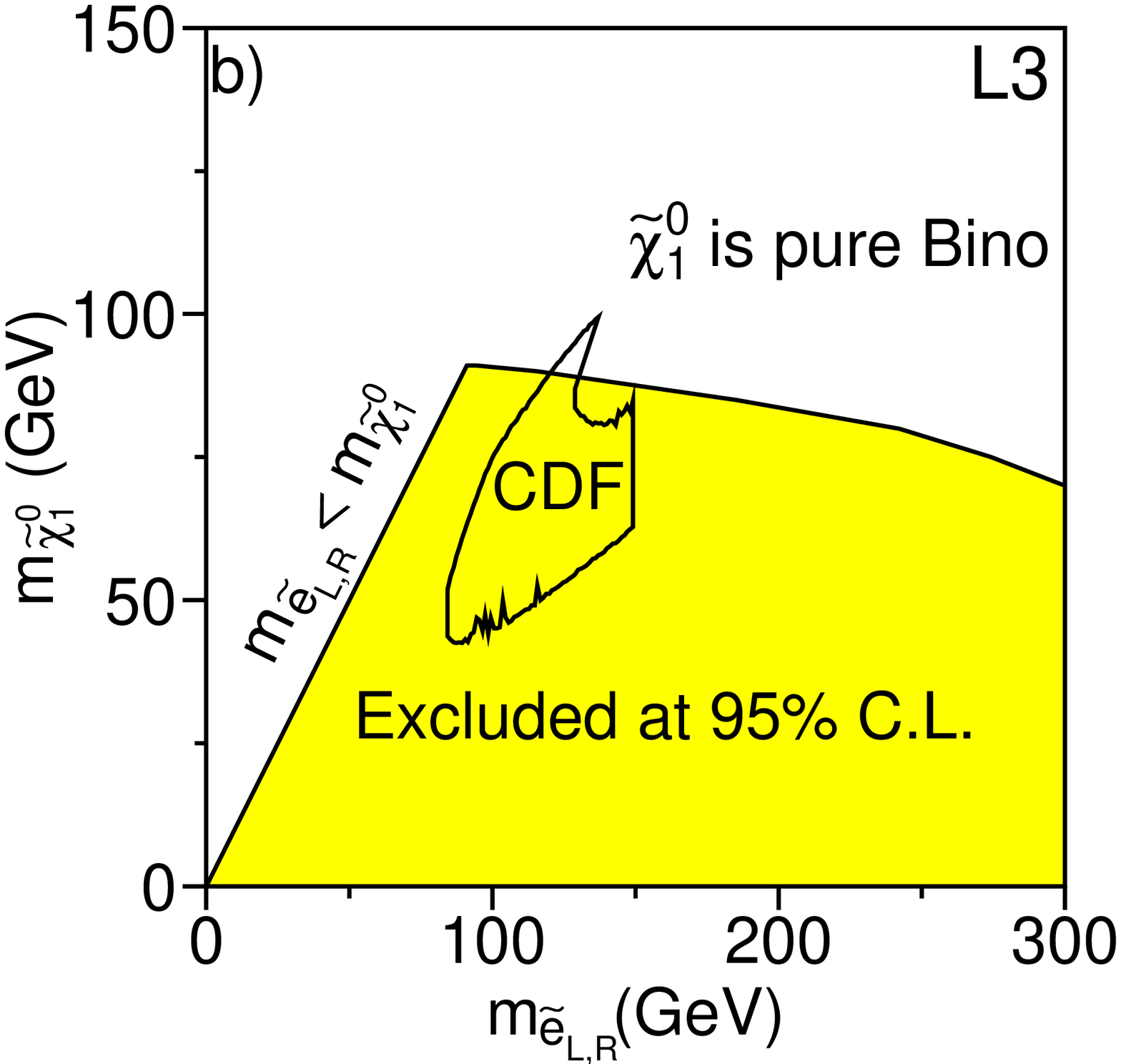}
    \icaption{\label{fig:n1ne1}(a) Upper limit on the cross section
      for \epem\ \ra\ \chinon\chinon\ \ra\
      \gravin\gravin\gam\gam. Theoretical predictions for two extreme
      cases of \chinon\  composition and for the most conservative
      GMSB prediction are also shown. (b) Excluded region for a pure
      bino neutralino model compared to the region consistent with the
      supersymmetric interpretation of the CDF event in the scalar
      electron scenario \protect\cite{ln96}.}
  \end{center}
\end{figure}

The selection described in this paper is devised for photons
originating from the interaction point. For a neutralino mean decay
length $d_{\chinon}$ larger than 1~cm the experimental sensitivity
drops. This problem arises only for peculiar situations in the light
gravitino scenario. The following limits in the gravitino LSP scenario
are derived under the assumption of $d_{\chinon} < 1$~cm. The cross
section limits for the process \epem\ \ra\ \chinon\chinon\ \ra\
\gravin\gravin\gam\gam\  are displayed in figure \ref{fig:n1ne1}(a)
versus neutralino mass. The efficiency varies between 17\% for \Mchi\
= 0.5 \gev{} and 62\% for \Mchi\ = 94 \gev. Theoretical predictions for
two extreme cases of neutralino content\footnote{For the higgsino
  case a 2\% photino component is required to ensure the decay into
  \gam\gravin.}
\cite{ln96}, which determines its coupling to the photon, are shown in
the same figure. For these cases of neutralino composition and for a
pure photino, we derive lower limits on the mass of the lightest
neutralino as listed in table \ref{tab:n1masslim}.
Figure \ref{fig:n1ne1}(b) shows the exclusion in the
$\chinon-\susylr{\e}$ mass plane derived with our data for a
neutralino being pure bino. The \e\e\gam\gam\  event observed by CDF
also has an interpretation in supersymmetric models with gravitino LSP
\cite{ln96}. Our analysis almost rules out this interpretation as
shown in figure \ref{fig:n1ne1}(b).

In minimal models with gauge-mediated SUSY breaking only five
parameters determine the sparticle sector of the theory \cite{gmsb}.
The parameters are $\Lambda$, the scale of SUSY breaking in the
messenger sector, $M_m$, the messenger mass scale, $N_m$, the number
of messenger fields, \tanb, the ratio of Higgs vacuum expectation
values.  In this model the absolute value of $\mu$, the higgsino mass
term, is fixed, however its sign is a free parameter. They have been
scanned to obtain neutralino masses, pair-production cross sections
and branching ratios for the decay to gravitino and photon. The scan
ranges on the individual parameters are \cite{neuprodgmsb}
   \( 10 \tev \leq \Lambda \leq 100 \tev,~
 \Lambda/0.9 \leq   M_m   \leq \Lambda/0.01,~
             N_m = 1\ldots 4,~
           1 \leq  \tanb  \leq 60,~
      \text{sign}\,\mu  = \pm 1 \).
The program {\tt ISASUSY} \cite{isasusy} has been used to calculate
sparticle masses and couplings from GMSB model parameters, and {\tt
SUSYGEN} to derive from these numbers the cross section for neutralino
pair-production including initial state radiation. Assuming a
neutralino NLSP scenario, the minimal cross section of \chinon\chinon\
production obtained within GMSB is shown in figure \ref{fig:n1ne1}(a),
which leads to a lower limit of
\begin{equation*}
  \Mchi > 88.2 \gev \, .
\end{equation*}

%
%
\section{Limits on Graviton Production}

Massive spin 2 gravitons propagating in 4+$\delta$ dimensions interact
with Standard Model particles with sizable strength in low scale
gravity models with extra dimensions \cite{qgrav1}. Gravitons produced
via \epem\ \ra\ \gam G lead to a single photon and missing energy
signature, since the graviton is not observed in the detector. The
reaction proceeds through $s$--channel photon exchange, $t$--channel
electron exchange and four-particle contact interaction \cite{extrad1}.

To convert the theoretical cross section of this process
\cite{extrad1} into an estimate on the number of events expected from
graviton production, the differential cross section in energy and
angle has been multiplied by efficiency and luminosity. The efficiency
is derived from \nnbar\gam(\gam) Monte Carlo simulation in a grid in
the $E_\gam - \cos\theta_\gam$ plane for $E_\gam > 4 \gev$. The
efficiency for \epem\ \ra\ \gam G within $E_\gam > 4 \gev$ and
$\cos\theta_\gam < 0.97$ is listed in table \ref{tab:extrad} for $2
\leq \delta \leq 10$.  The energy spectra shown in figure
\ref{fig:egamma}(b) and \ref{fig:recmegalo}(b) are used to derive
upper limits on the cross section.  They are listed in table
\ref{tab:extrad} together with the corresponding values for the energy
scale $M_D$.  These bounds improve on our previously published limits
\cite{graviton}.

%
%
\section{Acknowledgements}

We wish to express our gratitude to the CERN accelerator division for
the excellent performance of the LEP machine. We acknowledge the
effort of the engineers and technicians who have participated in the
construction and maintenance of this experiment.
\newpage

%
%
\bibliographystyle{l3style}

%
%
\newpage
\typeout{   }     
\typeout{Using author list for paper 186 -?}
\typeout{$Modified: Fri Sep 10 08:43:14 1999 by clare $}
\typeout{!!!!  This should only be used with document option a4p!!!!}
\typeout{   }
%
%
%
%
%
%

\newcount\tutecount  \tutecount=0
\def\tutenum#1{\global\advance\tutecount by 1 \xdef#1{\the\tutecount}}
\def\tute#1{$^{#1}$}
\tutenum\aachen            
\tutenum\nikhef            
\tutenum\mich              
\tutenum\lapp              
\tutenum\basel             
\tutenum\lsu               
\tutenum\beijing           
\tutenum\berlin            
\tutenum\bologna           
\tutenum\tata              
\tutenum\ne                
\tutenum\bucharest         
\tutenum\budapest          
\tutenum\mit               
\tutenum\debrecen          
\tutenum\florence          
\tutenum\cern              
\tutenum\wl                
\tutenum\geneva            
\tutenum\hefei             
\tutenum\seft              
\tutenum\lausanne          
\tutenum\lecce             
\tutenum\lyon              
\tutenum\madrid            
\tutenum\milan             
\tutenum\moscow            
\tutenum\naples            
\tutenum\cyprus            
\tutenum\nymegen           
\tutenum\caltech           
\tutenum\perugia           
\tutenum\cmu               
\tutenum\prince            
\tutenum\rome              
\tutenum\peters            
\tutenum\salerno           
\tutenum\ucsd              
\tutenum\santiago          
\tutenum\sofia             
\tutenum\korea             
\tutenum\alabama           
\tutenum\utrecht           
\tutenum\purdue            
\tutenum\psinst            
\tutenum\zeuthen           
\tutenum\eth               
\tutenum\hamburg           
\tutenum\taiwan            
\tutenum\tsinghua          
{
\parskip=0pt
\noindent
{\bf The L3 Collaboration:}
\ifx\selectfont\undefined
 \baselineskip=10.8pt
 \baselineskip\baselinestretch\baselineskip
 \normalbaselineskip\baselineskip
 \ixpt
\else
 \fontsize{9}{10.8pt}\selectfont
\fi
\medskip
\tolerance=10000
\hbadness=5000
\raggedright
\hsize=162truemm\hoffset=0mm
\def\r{\rlap,}
\noindent

M.Acciarri\r\tute\milan\
P.Achard\r\tute\geneva\ 
O.Adriani\r\tute{\florence}\ 
M.Aguilar-Benitez\r\tute\madrid\ 
J.Alcaraz\r\tute\madrid\ 
G.Alemanni\r\tute\lausanne\
J.Allaby\r\tute\cern\
A.Aloisio\r\tute\naples\ 
M.G.Alviggi\r\tute\naples\
G.Ambrosi\r\tute\geneva\
% ...
\newpage
\begin{list}{A}{\itemsep=0pt plus 0pt minus 0pt\parsep=0pt plus 0pt minus 0pt
                \topsep=0pt plus 0pt minus 0pt}
\item[\aachen]
 I. Physikalisches Institut, RWTH, D-52056 Aachen, FRG$^{\S}$\\
 III. Physikalisches Institut, RWTH, D-52056 Aachen, FRG$^{\S}$
\item[\nikhef] National Institute for High Energy Physics, NIKHEF, 
     and University of Amsterdam, NL-1009 DB Amsterdam, The Netherlands
\item[\mich] University of Michigan, Ann Arbor, MI 48109, USA
\item[\lapp] Laboratoire d'Annecy-le-Vieux de Physique des Particules, 
     LAPP,IN2P3-CNRS, BP 110, F-74941 Annecy-le-Vieux CEDEX, France
\item[\basel] Institute of Physics, University of Basel, CH-4056 Basel,
     Switzerland
\item[\lsu] Louisiana State University, Baton Rouge, LA 70803, USA
\item[\beijing] Institute of High Energy Physics, IHEP, 
  100039 Beijing, China$^{\triangle}$ 
\item[\berlin] Humboldt University, D-10099 Berlin, FRG$^{\S}$
\item[\bologna] University of Bologna and INFN-Sezione di Bologna, 
     I-40126 Bologna, Italy
\item[\tata] Tata Institute of Fundamental Research, Bombay 400 005, India
\item[\ne] Northeastern University, Boston, MA 02115, USA
\item[\bucharest] Institute of Atomic Physics and University of Bucharest,
     R-76900 Bucharest, Romania
\item[\budapest] Central Research Institute for Physics of the 
     Hungarian Academy of Sciences, H-1525 Budapest 114, Hungary$^{\ddag}$
\item[\mit] Massachusetts Institute of Technology, Cambridge, MA 02139, USA
\item[\debrecen] Lajos Kossuth University-ATOMKI, H-4010 Debrecen, Hungary$^\P$
\item[\florence] INFN Sezione di Firenze and University of Florence, 
     I-50125 Florence, Italy
\item[\cern] European Laboratory for Particle Physics, CERN, 
     CH-1211 Geneva 23, Switzerland
\item[\wl] World Laboratory, FBLJA  Project, CH-1211 Geneva 23, Switzerland
\item[\geneva] University of Geneva, CH-1211 Geneva 4, Switzerland
\item[\hefei] Chinese University of Science and Technology, USTC,
      Hefei, Anhui 230 029, China$^{\triangle}$
\item[\seft] SEFT, Research Institute for High Energy Physics, P.O. Box 9,
      SF-00014 Helsinki, Finland
\item[\lausanne] University of Lausanne, CH-1015 Lausanne, Switzerland
\item[\lecce] INFN-Sezione di Lecce and Universit\'a Degli Studi di Lecce,
     I-73100 Lecce, Italy
\item[\lyon] Institut de Physique Nucl\'eaire de Lyon, 
     IN2P3-CNRS,Universit\'e Claude Bernard, 
     F-69622 Villeurbanne, France
\item[\madrid] Centro de Investigaciones Energ{\'e}ticas, 
     Medioambientales y Tecnolog{\'\i}cas, CIEMAT, E-28040 Madrid,
     Spain${\flat}$ 
\item[\milan] INFN-Sezione di Milano, I-20133 Milan, Italy
\item[\moscow] Institute of Theoretical and Experimental Physics, ITEP, 
     Moscow, Russia
\item[\naples] INFN-Sezione di Napoli and University of Naples, 
     I-80125 Naples, Italy
\item[\cyprus] Department of Natural Sciences, University of Cyprus,
     Nicosia, Cyprus
\item[\nymegen] University of Nijmegen and NIKHEF, 
     NL-6525 ED Nijmegen, The Netherlands
\item[\caltech] California Institute of Technology, Pasadena, CA 91125, USA
\item[\perugia] INFN-Sezione di Perugia and Universit\'a Degli 
     Studi di Perugia, I-06100 Perugia, Italy   
\item[\cmu] Carnegie Mellon University, Pittsburgh, PA 15213, USA
\item[\prince] Princeton University, Princeton, NJ 08544, USA
\item[\rome] INFN-Sezione di Roma and University of Rome, ``La Sapienza",
     I-00185 Rome, Italy
\item[\peters] Nuclear Physics Institute, St. Petersburg, Russia
\item[\salerno] University and INFN, Salerno, I-84100 Salerno, Italy
\item[\ucsd] University of California, San Diego, CA 92093, USA
\item[\santiago] Dept. de Fisica de Particulas Elementales, Univ. de Santiago,
     E-15706 Santiago de Compostela, Spain
\item[\sofia] Bulgarian Academy of Sciences, Central Lab.~of 
     Mechatronics and Instrumentation, BU-1113 Sofia, Bulgaria
\item[\korea] Center for High Energy Physics, Adv.~Inst.~of Sciences
     and Technology, 305-701 Taejon,~Republic~of~{Korea}
\item[\alabama] University of Alabama, Tuscaloosa, AL 35486, USA
\item[\utrecht] Utrecht University and NIKHEF, NL-3584 CB Utrecht, 
     The Netherlands
\item[\purdue] Purdue University, West Lafayette, IN 47907, USA
\item[\psinst] Paul Scherrer Institut, PSI, CH-5232 Villigen, Switzerland
\item[\zeuthen] DESY, D-15738 Zeuthen, 
     FRG
\item[\eth] Eidgen\"ossische Technische Hochschule, ETH Z\"urich,
     CH-8093 Z\"urich, Switzerland
\item[\hamburg] University of Hamburg, D-22761 Hamburg, FRG
\item[\taiwan] National Central University, Chung-Li, Taiwan, China
\item[\tsinghua] Department of Physics, National Tsing Hua University,
      Taiwan, China
\item[\S]  Supported by the German Bundesministerium 
        f\"ur Bildung, Wissenschaft, Forschung und Technologie
\item[\ddag] Supported by the Hungarian OTKA fund under contract
numbers T019181, F023259 and T024011.
\item[\P] Also supported by the Hungarian OTKA fund under contract
  numbers T22238 and T026178.
\item[$\flat$] Supported also by the Comisi\'on Interministerial de Ciencia y 
        Tecnolog{\'\i}a.
\item[$\sharp$] Also supported by CONICET and Universidad Nacional de La Plata,
        CC 67, 1900 La Plata, Argentina.
\item[$\diamondsuit$] Also supported by Panjab University, Chandigarh-160014, 
        India.
\item[$\triangle$] Supported by the National Natural Science
  Foundation of China.
\item[\dag] Deceased.
\end{list}
}
\vfill






%
%

\begin{table}[H]
  \begin{center}
    \begin{tabular}{|l|r|r|r|r|r|} \hline
      & \multicolumn{3}{c|}{$E_\gam>5 \gev$} &
      \multicolumn{1}{c|}{$E_\gam>1.3 \gev$} &
      \multicolumn{1}{c|}{$E_{\gam_1}>5 \gev$} \\ \cline{2-4}
      & \multicolumn{1}{c|}{Total} & \multicolumn{1}{c|}{Barrel} &
      \multicolumn{1}{c|}{Endcaps} & 
      \multicolumn{1}{c|}{$E_\gam<5 \gev$} &
      \multicolumn{1}{c|}{$E_{\gam_2}>1 \gev$} \\ \hline\hline
      Data & 572 & 297 & 275 & 395 & 21 \\ \hline\hline
      \nnbar\gam(\gam) & 567.3 & 288.9 & 278.4 & 48.7 & 35.5 \\ \hline
      \epem\  background & 6.5 & 2.2 & 4.3 & 358.5 & 0.7 \\ \hline
      Cosmic background & 3.1 & 1.1 & 2.0 & 3.6 & 0 \\ \hline\hline
      Total expectation & 576.9 & 292.2 & 284.7 & 410.8 & 36.2 \\
      \hline
    \end{tabular}
    \icaption{\label{tab:sgamsel}Number of events selected in data,
      Monte Carlo predictions for processes from \epem\  collisions
      and contamination of cosmic ray background in the indicated
      kinematic regions.}
  \end{center}
\end{table}

\vspace*{-2mm}

\begin{table}[h]
  \begin{center}
    \begin{tabular}{|c|c|} \hline
      \rts\ (\GeV) & $N_\nu$ \\ \hline\hline
      130.1 & 2.63 $\pm$ 0.40 $\pm$ 0.10 \\ \hline
      136.1 & 2.98 $\pm$ 0.49 $\pm$ 0.14 \\ \hline
      161.3 & 3.68 $\pm$ 0.53 $\pm$ 0.09 \\ \hline
      172.1 & 4.24 $\pm$ 0.65 $\pm$ 0.09 \\ \hline
      182.7 & 3.13 $\pm$ 0.26 $\pm$ 0.05 \\ \hline
      188.6 & 2.94 $\pm$ 0.15 $\pm$ 0.04 \\ \hline\hline
      130 -- 189 & 3.05 $\pm$ 0.11 $\pm$ 0.04 \\ \hline
      88 -- 94 & 2.98 $\pm$ 0.07 $\pm$ 0.07 \\ \hline\hline
      Average & 3.011 $\pm$ 0.077 \\ \hline
    \end{tabular}
    \icaption{\label{tab:nonuav}Number of neutrino families measured
      from single photon events.}
  \end{center}
\end{table}

\vspace*{-2mm}

\begin{table}[h]
  \begin{center}
    \begin{tabular}{|c|c|c|} \hline
      \chinon\  content & $m_{\susylr{\e}}$ (\GeV) &
      $\Mchi^\mathrm{lim}$ (\GeV) \\[1mm] \hline\hline 
      Bino & 150 & 87.9 \\ \hline
      Bino & 100 & 90.8 \\ \hline
      Photino & 150 & 88.3 \\ \hline
      Photino & 100 & 91.1 \\ \hline
      Higgsino & --- & 89.0 \\ \hline
    \end{tabular}
    \icaption{\label{tab:n1masslim}Neutralino mass limits for several
      neutralino compositions and selectron masses.}
  \end{center}
\end{table}

\vspace*{-2mm}

\begin{table}[h]
  \begin{center}
    \begin{tabular}{|l||c|c|c|c|c|c|c|c|c|c|} \hline
      $\delta$ & 2 & 3 & 4 & 5 & 6 & 7 & 8 & 9 & 10 \\ \hline
      $\epsilon$ (\%) & 42.8 & 40.7 & 38.9 & 37.6 & 36.5 & 35.5 & 34.7 &
      34.0 & 33.4 \\ \hline
      $\sigma_{\gam\mathrm{G}}^\mathrm{lim}$ (pb) & 0.638 & 0.646 &
      0.651 & 0.658 & 0.664 & 0.670 & 0.674 & 0.678 & 0.680 \\ \hline
      $M_D$ (\GeV) & 1018 & 812 & 674 & 577 & 506 & 453 & 411 & 377 &
      349 \\ \hline
    \end{tabular}
    \icaption{\label{tab:extrad}Selection efficiency $\epsilon$ for
      \epem\ \ra\ \gam $\mathrm{G}$, upper cross section limit and
      lower limit on the energy scale $M_D$ as a function of the
      number of extra dimensions $\delta$.}
  \end{center}
\end{table}

\clearpage
\newpage
%
%

\clearpage
\newpage

\clearpage
\newpage

\clearpage
\newpage

\clearpage
\newpage

\clearpage
\newpage

\clearpage
\newpage

\clearpage
\newpage

\end{document}